\documentstyle[graphicx,float]{article}

\begin{document}
\title{Two-particle superposition effects in two-slit interference}
\date{}
\author{Pedro Sancho \\ Centro de L\'aseres Pulsados CLPU \\ Parque Cient\'{\i}fico, 37085 Villamayor, Salamanca, Spain}
\maketitle
\begin{abstract}
Entanglement can modify the interference patterns of multi-particle systems. We analyse, using  the path integral formalism, a novel example of multi-particle interference and some unexplored aspects of this phenomenon by considering the two-slit arrangement with two distinguishable particles in a superposition state. The two components of entanglement, multiplicity of multi-particle terms and their coherence, can be studied separately by comparing the patterns of product, mixed and superposition states. This suggests a scheme to detect multi-particle superposition in some favourable cases. Moreover, the dependence of the effect on the entanglement degree, measured by the Schmidt number, is not a monotone.  
\end{abstract}

\section{Introduction}
Multi-particle superposition leads to multi-particle interference \cite{Ze1,Ze2}. This phenomenon, induced by entanglement, has been extensively studied, and many examples of this behaviour have been presented in the literature. Many of these ideas have been experimentally analysed, most of them taking advantage of the correlations between photons generated in parametric down-conversion \cite{wal}. There are also examples of multi-particle interference of massive particles, as the diffraction of entangled atoms by light gratings via the Kapitza-Dirac effect \cite{yap}. A recurrent topic in the field is the existence of complementarity relations between one- and two-particle interference. For instance, the two-particle two-two-slit arrangement \cite{tts} provides an experimental demonstration of these relations. 

The two-slit arrangement is an archetypical tool to study one- and multi-particle interference. It allows for rather simple analytical evaluations of interference patterns, making it a powerful theoretical laboratory. Many gedanken experiments have been carried out this way. From the experimental point of view, nowadays it is even possible to carry out them with a single electron passing the slits in every repetition of the experiment \cite{bat}. In this paper we want to use this arrangement to show that it provides a new instance of multi-particle interference and to study some aspects of the phenomenon previously not considered in the literature. Many studies use a pair of entangled particles incident on two spatially separated two-slit devices (each particle interacts with a different device) \cite{tts,bra}. Here, in contrast, the set-up contains only one two-slit device and both particles are incident on it. A similar scheme has been previously considered to analyse interference of massive identical particles \cite{yjp} .  In the case of light several experimental studies involving several photons but only one two-slit (or multiple slits) device have been carried out (see references \cite{1} to\cite{5}).  We evaluate the detection patterns using the path integral formalism of Feynman \cite{fey}.  The explicit calculation of the detection patterns was carried out in \cite{yjp} and we do not need to repeat here the evaluation. We only have to adapt these results to the case of distinguishable particles. We shall find multi-particle superposition deviations with respect to the interference patterns found in non-entangled  systems. 

Entanglement is the phenomenon of multi-particle superposition. It has two components, the presence of various multi-particle terms and the coherence between them. Via the direct comparison between the patterns of product and superposition states it is not possible to separate the two contributions. We shall find that if we also compare with the pattern of mixed states it will be possible to separately analyse them and to study their interplay. As a by-product of this approach we show that for some favourable values of the parameters of the arrangement, a two-step scheme comparing the three patterns provides a method  to detect multi-particle superposition in situations where otherwise the differences between the patterns of product and superposition states are very small.
 
The second aspect of multi-particle interference we shall consider in the paper is the dependence on the entanglement degree. Do the modifications associated with the multi-particle superposition increase with the entanglement degree?  An entanglement measure well-suited to study continuous variables is the Schmidt number \cite{sch,sc1}. The main result of our analysis is that the modifications generated by the superposition are not a monotone of the Schmidt number.  As a by-product of our analysis we consider another interesting aspect in our problem, the fact that the components of the multi-particle superposition are not orthogonal. The studies of the entanglement of non-orthogonal states in the literature are scarce in comparison with those of orthogonal ones, but could be on the basis of some interesting applications  (see, for instance, \cite{mun,otr}) . The non-orthogonality leads to a double dependence of the Schmidt number, on the coefficients of the superposition and on the overlap of the two components, measuring the last one the non-orthogonality of the system.    

\section{The arrangement}

First of all, we briefly describe the arrangement (see Fig. 1). A source prepares pairs of distinguishable particles in superposition states
\begin{equation}
\Psi ({\bf x},{\bf y})=N(a\psi ({\bf x})\phi ({\bf y})+b\varphi ({\bf x}) \chi ({\bf y}))
\end{equation}
The wave functions $\psi$ and $\varphi$ and the spatial coordinate ${\bf x}$ make reference to one of the particles, whereas $\phi$, $\chi$ and ${\bf y}$ do to the other. $a$ and $b$ are the coefficients of the superposition and obey the relation $|a|^2+|b|^2=1$. All the one-particle wave functions are normalized. On the other hand, the full wave function normalization factor is
\begin{equation}
N=(1+2Re(a^*b<\psi|\varphi><\phi|\chi>))^{-1/2}
\end{equation}
with $<\psi|\varphi>$ and $<\phi|\chi>$ the scalar product of the one-particle wave functions, which must be non-orthogonal, $<\psi|\varphi> \neq 0$ and $<\phi|\chi> \neq 0$, in order to travel towards the same two-slit device.
\begin{figure}[H]
\centering
\includegraphics[width=8cm,height=4cm]{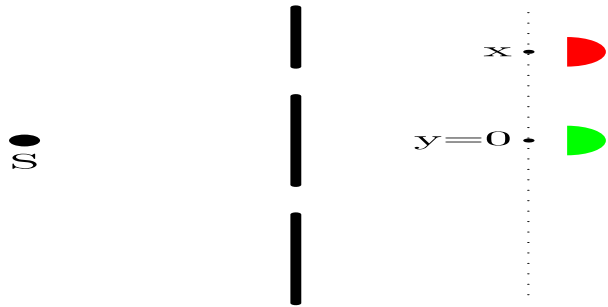}
\caption{Schematic representation of the arrangement. The source S emits pairs of entangled particles. After the slits we measure the simultaneous arrival of particles along the dotted line, parametrized by the coordinate $x$ for one particle and $y$ for the other. The green detector is fixed at $y=0$, whereas the red one can be displaced along the line.}
\end{figure}
An example of a potential source of entangled atoms for this type of problems was discussed in \cite{yjp}. It is based on atomic traps that release particles in a controlled way. In our case the traps should contain atoms of two distinguishable types. The photo-dissociation of molecules has also been used for the preparation of pairs of atoms in non-separable states \cite{jap,bel}.  A method to generate the superposition of photo-dissociated atoms in a controlled way was proposed in \cite{yyy}. 

After leaving the source the particles impinge on the two slits. The width of both slits is $2b_s$. After the slits we place detectors to measure the simultaneous arrival of particles at different points.  If the detectors are in a plane parallel to that containing the slits, the problem can be simplified to an one-dimensional one. The relevant dimension is that of the line containing the two slits or, equivalently, that of the detectors. The coordinates corresponding to that dimension are denoted by $x$ and $y$. With this notation the source is placed at $x=0$ and the middle points of the slits at $\pm x_0$.

In order to evaluate the evolution of the wave function we must specify its initial form. We assume the one-particle states to be multi-mode ones, and as usual we take them in the form \cite{ada}
\begin{equation}
\psi (x.t) =(2\pi)^{-1} \int dk f(k) \exp (i(kx-k^2\hbar t/2m))
\end{equation}
with the mode distribution given by
\begin{equation}
f(k)=\frac{(4\pi)^{1/4}}{\sigma ^{1/2}} \exp(-k^2/2\sigma^2)
\end{equation}
Similar expressions are valid for the rest of one-particle states using different coefficients ($\bar{\sigma}$ for $\varphi$, $\xi$ for $\phi$, and $\bar{\xi}$ for $\chi$).

The calculation of the wave function can be done using the path integral approach. In this framework the passage through the slits can be modelled via the Gaussian slit approximation, where the finite range of integration associated with the slits is replaced by an infinite one, but weighted  by the Gaussian function $\exp(-(x-x_0)^2/2b_s^2)$ \cite{fey}. This way the complex Fresnel functions 
arising in the exact evaluation are replaced by Gaussian ones, simplifying the calculations.

The explicit evaluation of the wave function was carried out in \cite{yjp} for identical particles, and the details of the calculations can be seen in that reference. It is immediate to translate the results to the case of distinguishable particles. By the sake of completeness they are included in the Appendix 1.

\section{Interference patterns}

The interference patterns can be derived from the wave functions in the Appendix 1. The final wave function is obtained from $\Psi$ with the replacements $\psi \rightarrow \psi _A + \psi _B$, $\cdots$: 
\begin{eqnarray}
\Phi (x,y)=Na(\psi_A ( x)+\psi_B(x) )(\phi_A (y)+\phi_B(y))+ \nonumber \\
Nb(\varphi_A (x)+\varphi_B(x))( \chi_A (y) +\chi_B(y))
\end{eqnarray}
where the subscripts $A$ and $B$ refer to the two slits.  The evolution in the region of the slits is not unitary because the particle interacts with the solid elements surrounding the slits and can be absorbed. In the experiment the number of particles passing beyond the slits is smaller than the number of particles generated at the source. In the path integral formalism this non-unitarity is represented by the paths blocked by the solid part of the arrangement. Thus, we must normalize again the final wave function
\begin{equation}
\bar{\Phi}=\frac{\Phi (x,y)}{\left(\int dx \int dy |\Phi (x,y)|^2 \right)^{1/2}}
\end{equation}
The normalization factor can be calculated analytically, but it leads to a rather lengthy and cumbersome expression. Thus, we perform the normalization numerically.
 \begin{figure}[H]
\centering
\begin{tabular}{cc}
\includegraphics[width=6cm,height=7cm]{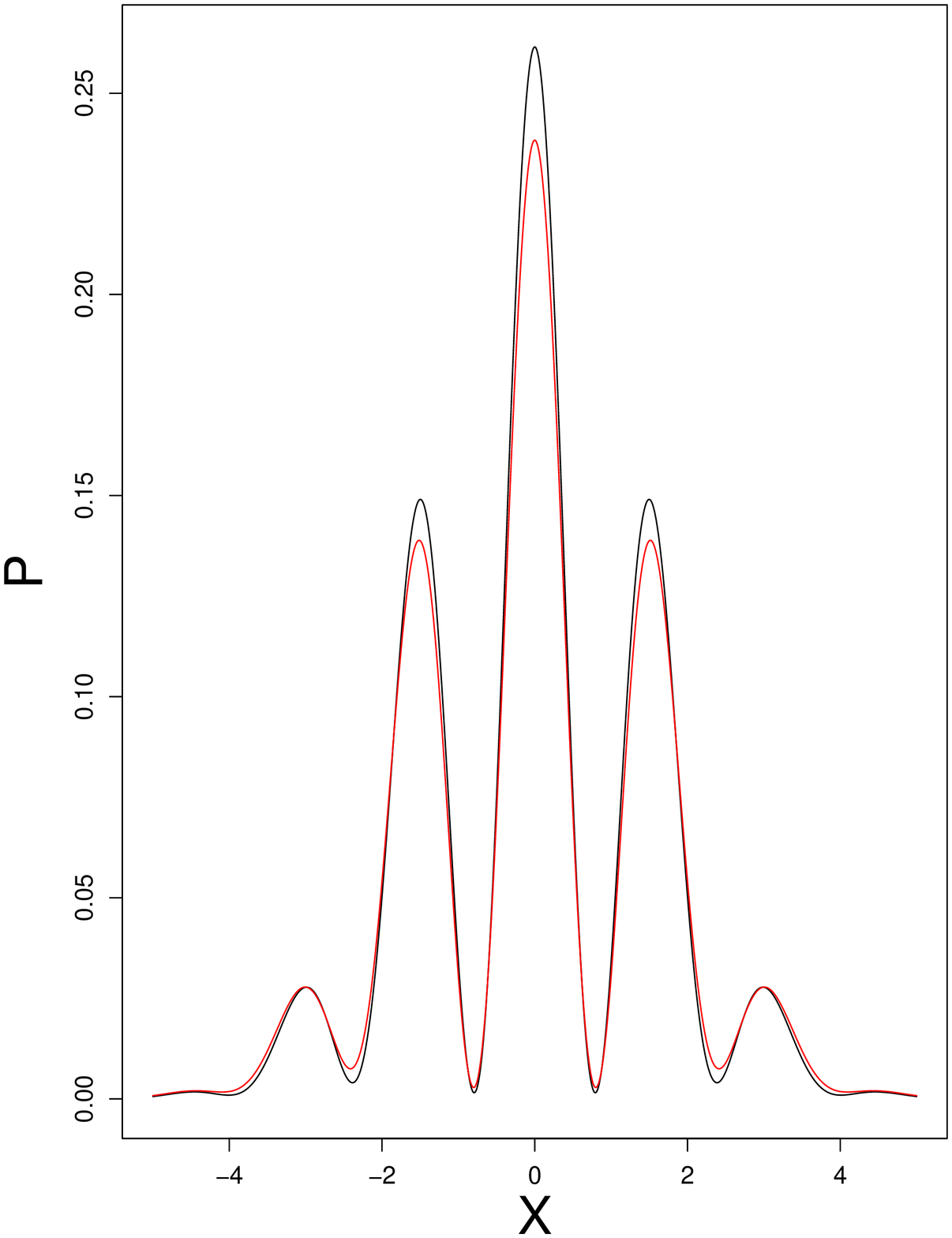}&
\includegraphics[width=6cm,height=7cm]{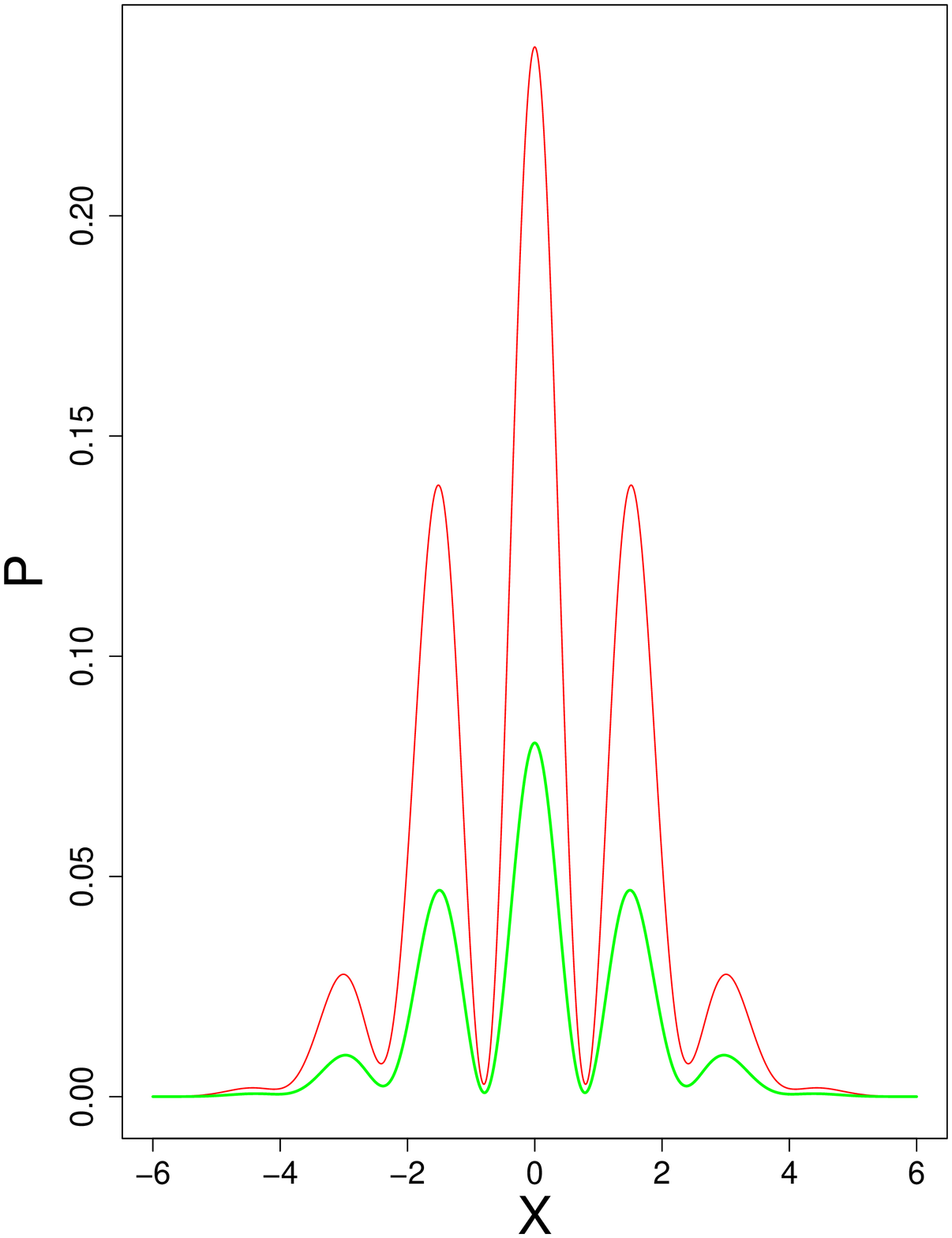}
\end{tabular}
\caption{Representation of the probability density of simultaneous double detection, $P$ in $\mu m^{-2}$, versus the position of the movable detector, $x$ in $\mu m$. The left-side figure contains the product state ($a=1$, black curve) and an entangled state ($a=0.3$, red line). The right-side figure compares the same entangled state with the corresponding mixture (green curve).}
\end{figure}
Next, we represent graphically the detection patterns. As usual in this type of problem we consider coincidence detections at two points $x$ and $y$, and we fix the position of one of them, for instance $y=0$. In Appendix 2 we analyse what happens when we change the fixed point.  As signalled before, we have evaluated numerically  the normalization factor for each graphic.  The range of values of the parameters to be used in the representation has been discussed in \cite{yjp}: $b_s=0.1 \mu m$, $x_0=0.4 \mu m$, $\hbar(t-t_s)/m_1=0.2 \mu m^2$, $\hbar t_s/m_1=0.33 \mu m^2$ and $\sigma =1 \mu m^{-1}$. Similarly, we take  $0.9 \hbar t_s/m_2=\hbar t_s/m_1$, $0.9 \hbar (t-t_s)/m_2=\hbar (t-t_s)/m_1$, $\bar{\sigma}=6$ and, moreover, we assume $\xi = \sigma$ and $\bar{\xi} = \bar{\sigma}$.

We rewrite the normalized final wave function as $\bar{\Phi}={\cal N}(a\Phi _a +b\Phi _b)$, with $\Phi _a=(\psi _A+\psi_B)(\phi_A+\phi_B)$, an obvious similar expression for $\Phi_b$, and ${\cal N}=N/\left(\int dx \int dy |\Phi (x,y)|^2 \right)^{1/2}$. The probabilities for superposition ($\bar{\Phi}$), mixture (mixed state with density matrix $|a|^2{\cal N}_a^2|\Phi _a><\Phi _a|+ |b|^2{\cal N}_b^2|\Phi _b><\Phi _b|$) and product state ($\Phi _a$) are $P=|\bar{\Phi}|^2$, $P_{mix}=|a|^2{\cal N}_a^2|\Phi_a|^2+|b|^2{\cal N}_b^2|\Phi_b|^2$ and $P_{pro}^a={\cal N}_a^2|\Phi_a|^2$, where ${\cal N}_a={\cal N}(a=1)$ and ${\cal N}_b={\cal N}(b=1)$ are the normalization coefficients of $\Phi_a$ and $\Phi _b$. Note that the states $\Phi _a$ and $\Phi_b$,  as $\Phi$, must be normalized again after passing the slits because of the non-unitary evolution.
We first compare the probability density of simultaneous detection of a product state ($a=1$) with that of an entangled one ($a=0.3$). We see that the presence of multi-particle superposition leads to different detection patterns. Both curves have similar forms, but the peaks reach different values. For other values of $a$ and the rest of parameters we find a similar behaviour. We also compare the detection patterns of the same entangled state with a mixture with the same weights, $0.3^2$ for the first term and $1-0.09$ for the second one. Now the differences between the entangled and non-entangled states are much sharper than in the previous case. The position of the peaks is the same for both curves, but the values are very different. The modifications associated with multi-particle superposition show larger values than the corresponding mixture. 

A common tool to compare interference figures is the visibility, $V=(P_{max}-P_{min})/(P_{max}+P_{min})$, with the subscripts denoting the maxima and minima values of $P$. In our example the values for the superposition, mixture and product state for the three central peaks (the values for only the central peak are very similar) are $V=0.94$, $V_{mix}=0.92$ and $V_{pro}=0.98$. The visibilities are very similar due to the very small values of the minima (when all the pattern is taken into account we have $V=V_{mix}=V_{pro}=1$ because the minimum value is null). The visibility does not capture the large differences between the superposition and the mixture, which are more adequately represented by the maximum values of the central peaks. Consequently, in this case the visibility is not a good measure of the loss of coherence in the transition from a superposition to a mixture. 

It is also interesting to study how these differences depend on how much the non pure state is mixed. The measure of the degree of mixedness is the purity, which in our case reads $|a|^4+|b|^4+2|a|^2|b|^2|<\psi|\varphi>|^2|<\phi|\chi>|^2$.  With the values used in Fig. 2 and the expression of the scalar products that will be derived later (Eq. (\ref{eq:cat})), we obtain $0.85$ for the purity. We have that even for these intermediate/large values of the purity we can have large deviations. 

The behaviour of the above patterns can be explained analysing in detail the structure of the detection probabilities. We want to understand the two most relevant characteristics of the patterns, (i) the shape of all the curves is the same (in the sense that the position of the maxima and minima of the different cases is almost equal), and (ii) the changes of the mixture and the product state with respect to the superposition are very different.  When we represent numerically ${\cal N}_a^2|\Phi_a|^2$ and ${\cal N}_b^2|\Phi_b|^2$ we can see that the position of the maxima and minima is almost identical, although the values of the maxima are different. This property can be easily justified remembering that the two components of the superposition are Gaussian packets picked around the same central value. As the form of the final patterns mainly depends on the central part of the packet, we expect to have very similar spatial distributions for the maxima and minima.  We have also numerically studied the term of interference in the superposition, $2{\cal N}^2Re(a^*b\Phi_a^*\Phi_b)$, that it is always much smaller than the other terms (and with the maxima and minima at different positions). Then, we can approximate $P \approx {\cal N}^2(|a|^2|\Phi_a|^2+|b|^2|\Phi_b|^2)$. Note that this approximation does not imply that the interference effects are negligible. They are also present in the normalization factor, which has a very different value of these of ${\cal N}_a$ and ${\cal N}_b$ appearing in $P_{mix}$. Taking into account the approximate form of $P$ and the equivalent shape of ${\cal N}_a^2|\Phi_a|^2$ and ${\cal N}_b^2|\Phi_b|^2$ we can justify the point (i). Using an input state with a different form of the wave packets would lead  in most cases to detection distributions with different shapes.

In order to study (ii), that is, the differences between the two relative distributions, we use their approximate forms 
\begin{equation}
P-P_{mix} \approx |a|^2({\cal N}^2-{\cal N}_a^2)|\Phi _a|^2+|b|^2({\cal N}^2-{\cal N}_b^2)|\Phi _b|^2
\end{equation} 
and
\begin{equation}
P-P_{pro}^a \approx (|a|^2{\cal N}^2-{\cal N}_a^2)|\Phi _a|^2+|b|^2{\cal N}^2|\Phi _b|^2
\end{equation} 
We study the case of $|a|^2$ small (in Fig. 2 is $|a|^2=0.09$). As ${\cal N}$ and ${\cal N}_a$ are coefficients of the same order of magnitude we have that $|a|^2{\cal N}^2 \ll {\cal N}_a^2$ and we can approximate the second expression as  $P-P_{pro}^a \approx |b|^2{\cal N}^2|\Phi _b|^2-{\cal N}_a^2|\Phi _a|^2$. We see that for small values of $|a|$ the two terms in the right hand side have opposite signs and their contributions are subtracted. On the other hand, as from the numerical calculations we have that ${\cal N}>{\cal N}_a$ and  ${\cal N}>{\cal N}_b$, the two terms in $P-P_{mix}$ have the same sign and their contributions are added. Then, for $|a|$ small we expect the differences induced by the superposition with respect to the product state to be smaller than in the case of the mixture. 

The difference between the relative patterns is not always so large. The difference between them is given by the function $D=(P-P_{mix})-(P-P_{pro}^a)$, which is a function of $x$. As $P_{mix}=|a|^2P_{pro}^a+|b|^2P_{pro}^b$ we have 
\begin{equation}
D=|b|^2(P_{pro}^a-P_{pro}^b)
\end{equation} 
Thus, the difference function is at every point $x$ a parabola, with null value at $|b|=0$ and that increases with $|b|$, reaching the maximum value $P_{pro}^a-P_{pro}^b$. The case represented in Fig. 2, $a=0.3$, corresponds to large differences between both relative patterns, but for small values of $|b|$ we have the opposite behaviour of very similar relative curves.  The difference function is independent of $P$ and only depends on $P_{pro}^a-P_{pro}^b$. This is an immediate consequence of the fact that we are comparing $P_{mix}$ and $P_{pro}^a$ with the same pattern $P$ and, consequently, the relative differences can only correspond to the differences between $P_{mix}$ and $P_{pro}^a$.

We conclude that the main characteristics of Fig. 2 can be explained invoking the particular form of the input state (fundamental to justify the similitude of the patterns shape) and the effects of superposition (necessary to account for the different intensities of the patterns). The manifestation of the superposition effects differs from textbook examples, where it corresponds to the addition of an interference term ($2{\cal N}^2Re(a^*b\Phi_a^*\Phi_b)$) to the direct ones (${\cal N}^2|a|^2|\Phi_a|^2$ and ${\cal N}^2|b|^2|\Phi_b|^2$). This addition in our arrangement is very small . In our case the superposition effects manifest through the normalization coefficient ${\cal N}$ and the differences with ${\cal N}_a$ and ${\cal N}_b$.

\section{Dependence on Schmidt's number}

In the second part of the paper we focus on the dependence of the effects discussed above on the entanglement degree. For continuous variables, an entanglement measure frequently used in the literature is the Schmidt number \cite{sch}. It is defined as
\begin{equation}
S=\frac{1}{Tr_x (\hat{\rho}_x^2)}=\frac{1}{Tr_y (\hat{\rho}_y^2)}
\end{equation}  
with $\hat{\rho}_x=Tr_y(|\Psi ><\Psi |)$ and $\hat{\rho}_y=Tr_x(|\Psi ><\Psi |)$ the two reduced density matrices. In the above equations $Tr _i$ denotes the trace with respect to the variable $i=x,y$. The state $|\Psi>$ can be expressed in the position representation as
\begin{equation}
|\Psi>= \int dx \int dy \Psi (x,y)|x>|y>
\end{equation}
Using this representation we can evaluate in a simple way the Schmidt number:
\begin{equation}
S^{-1}=\int dx \int dy \int dX \int dY \Psi ^*(x,y) \Psi (X,y) \Psi ^*(X,Y) \Psi (x,Y)
\end{equation}
At the initial time, $t=0$, which we identify with the time of the preparation of the state the one-particle wave functions are $\psi (x)=(\sigma ^{1/2}/\pi^{1/4}) \exp (-\sigma ^2 x^2/2), \cdots$ \cite{yjp}. Using, as in the graphical representation above, the conditions $\xi = \sigma$ and $\bar{\xi} = \bar{\sigma}$, the initial state is
\begin{equation}
\Psi (x,y)= Na\frac{\sigma }{\pi^{1/2}} e^{-\sigma ^2(x^2+y^2)/2} + Nb\frac{\bar{\sigma }}{\pi^{1/2}} e^{-\bar{\sigma} ^2(x^2+y^2)/2}
\end{equation}
The scalar products at the initial time, which we denote from now on by $\theta$, are
\begin{equation}
\theta =<\psi|\varphi>_{t=0} =<\phi|\chi>_{t=0}= \left(  \frac{2\sigma \bar{\sigma}}{\sigma^2+\bar{\sigma}^2} \right) ^{1/2}
\label{eq:cat}
\end{equation} 
Note that both scalar products are equal because at the initial time the wave functions are independent of the mass and of our choice for $\xi$ and $\bar{\xi}$. The normalization factor can be expressed as
\begin{equation}
N=\left( 1+ \frac{4\sigma \bar{\sigma}}{\sigma^2 + \bar{\sigma}^2} Re(a^*b)  \right) ^{-1/2}
\end{equation}
Finally, after a lengthy but straightforward calculation we have
\begin{equation}
S=\frac{(1+2ab\theta^2)^2}{a^4+b^4+4ab\theta^2+2a^2b^2\theta^2(2+\theta^2)}
\end{equation} 
where we have assumed the coefficients $a$ and $b$ to be real.

The Schmidt number depends on two factors, the coefficients of the superposition and the overlapping $\theta$ between the one-particle states of each particle. The first dependence is common to all the forms of entanglement, which is a multi-particle superposition phenomenon. The second one is only present in non-orthogonal systems, as the example considered here, where the one-particle states must overlap in order to reach the same slits. We remark that we refer to one-particle non-orthogonality, not to the spatial overlapping between the two particles.  
\begin{figure}[H]
\centering
\begin{tabular}{cc}
\includegraphics[width=6cm,height=5cm]{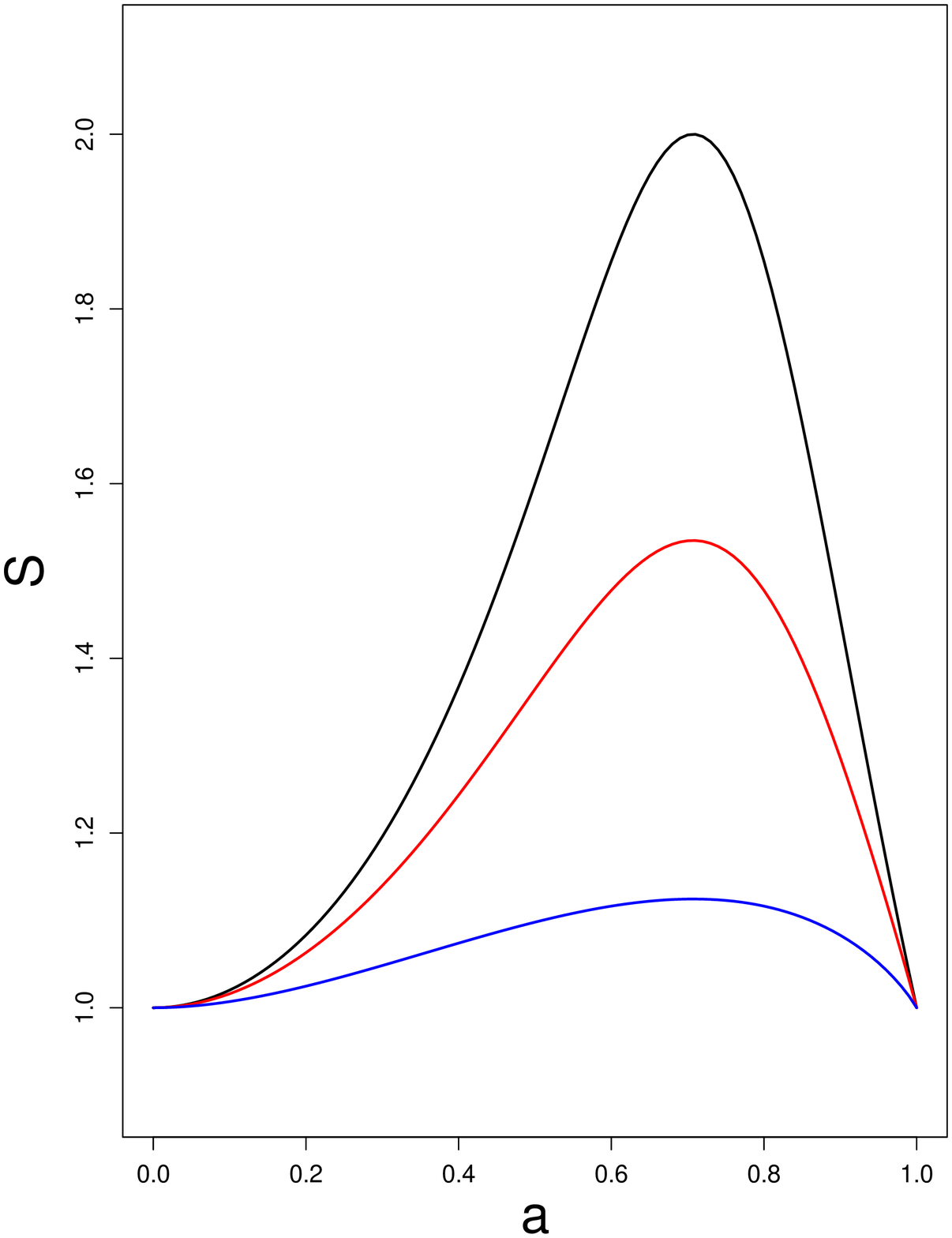}&
\includegraphics[width=6cm,height=5cm]{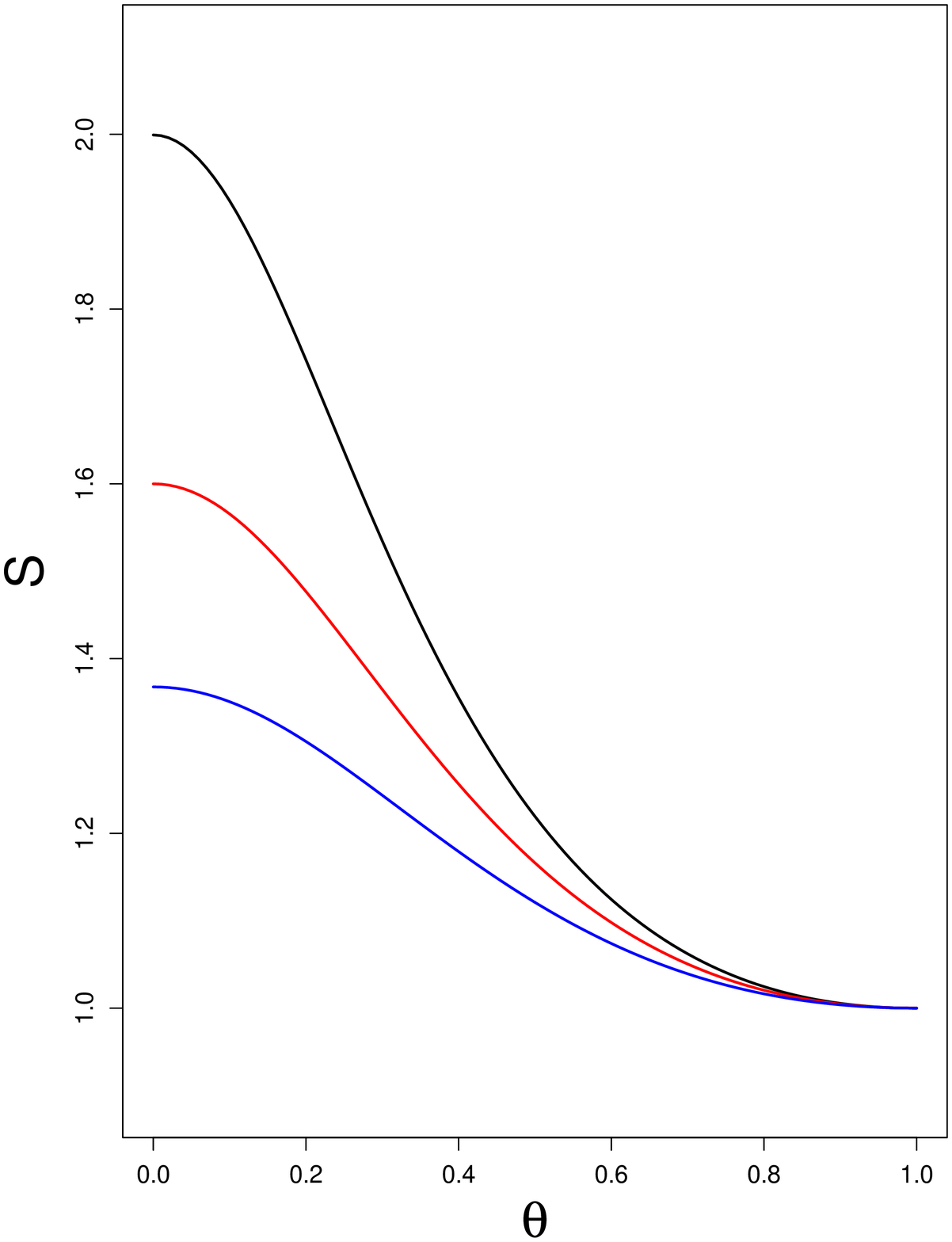}
\end{tabular}
\caption{Representation of the Schmidt number $S$ versus respectively the $a$ coefficient of the superposition (left-side graph), and the one-particle overlap $\theta$ (right-side graphic). The black, red and blue curves correspond respectively to the values $\theta =0,0.3,0.6$ in the first figure, and $a=0.7,0.5,0.4$ in the second one.}
\end{figure}
In order to see in a pictorial way these dependences we represent them in Fig. 3.  We consider first the dependence of $S$ on the superposition coefficient $a$ in Fig. 3 left. For $a=0$ and $a=1$ there is no superposition and we have $S=1$, representing the absence of entanglement. The three curves, corresponding to different values of one-particle overlap, reach their maximum values for $a=1/\sqrt 2$ ($a=b$ for real coefficients) but the values of these peaks are different. For any value of the superposition coefficient the Schmidt number decreases with increasing overlap. The non-overlapping one-particle states are the more entangled ones. On the other hand, when the overlap is close to one the state tends toward a product one and the entanglement is negligible.  With respect to the dependence of $S$ on the overlap $\theta$, represented in Fig. 3 right, we again see that always decreases for increasing values of the non-orthogonality degree. The rate of decrease grows with the value of $a$. 

The above results agree with the common understanding of multi-particle superposition. The interesting aspect emerges when we compare the modifications induced in the detection patterns with the entanglement degree. We graphically present this comparison in Fig. 4. The separation between the curves corresponding to superposition and that to a product state depends on the values of the coefficients of the superposition. For $a=0.3$ it is larger than for $a=0.7$, although as we can see in the right-side graph the Schmidt number is larger in the second case. We have found the same behaviour for other values of the parameters. On the other hand, when we consider the comparison with respect to a mixture, for some values the separation increases with the entanglement degree and for others we observe the opposite behaviour. We conclude that there is not an universal trend of the separations with respect to the Schmidt number. The effects associated with multi-particle superposition are not a monotone of the entanglement degree. 
\begin{figure}[H]
\centering
\begin{tabular}{cc}
\includegraphics[width=6cm,height=5cm]{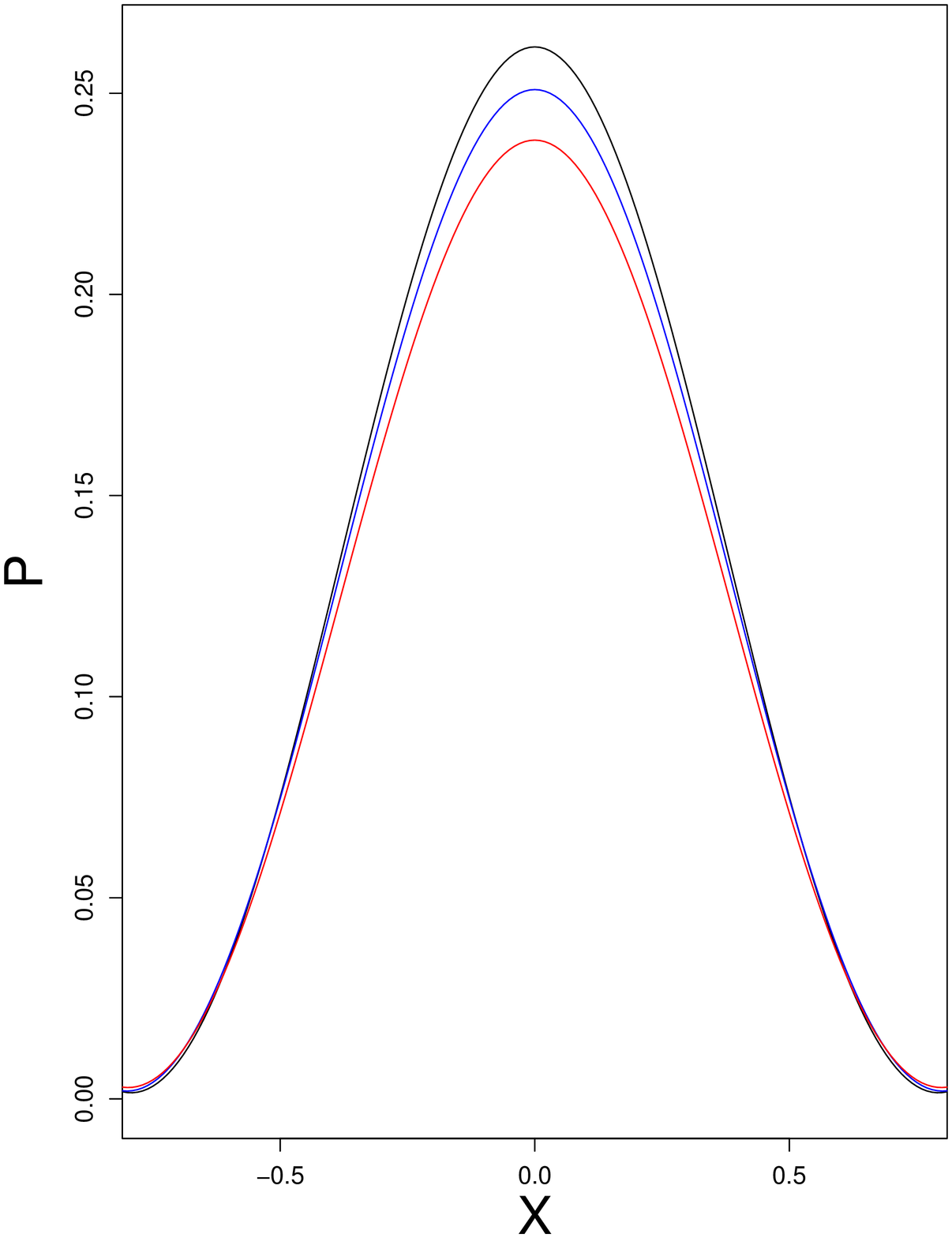}&
\includegraphics[width=6cm,height=5cm]{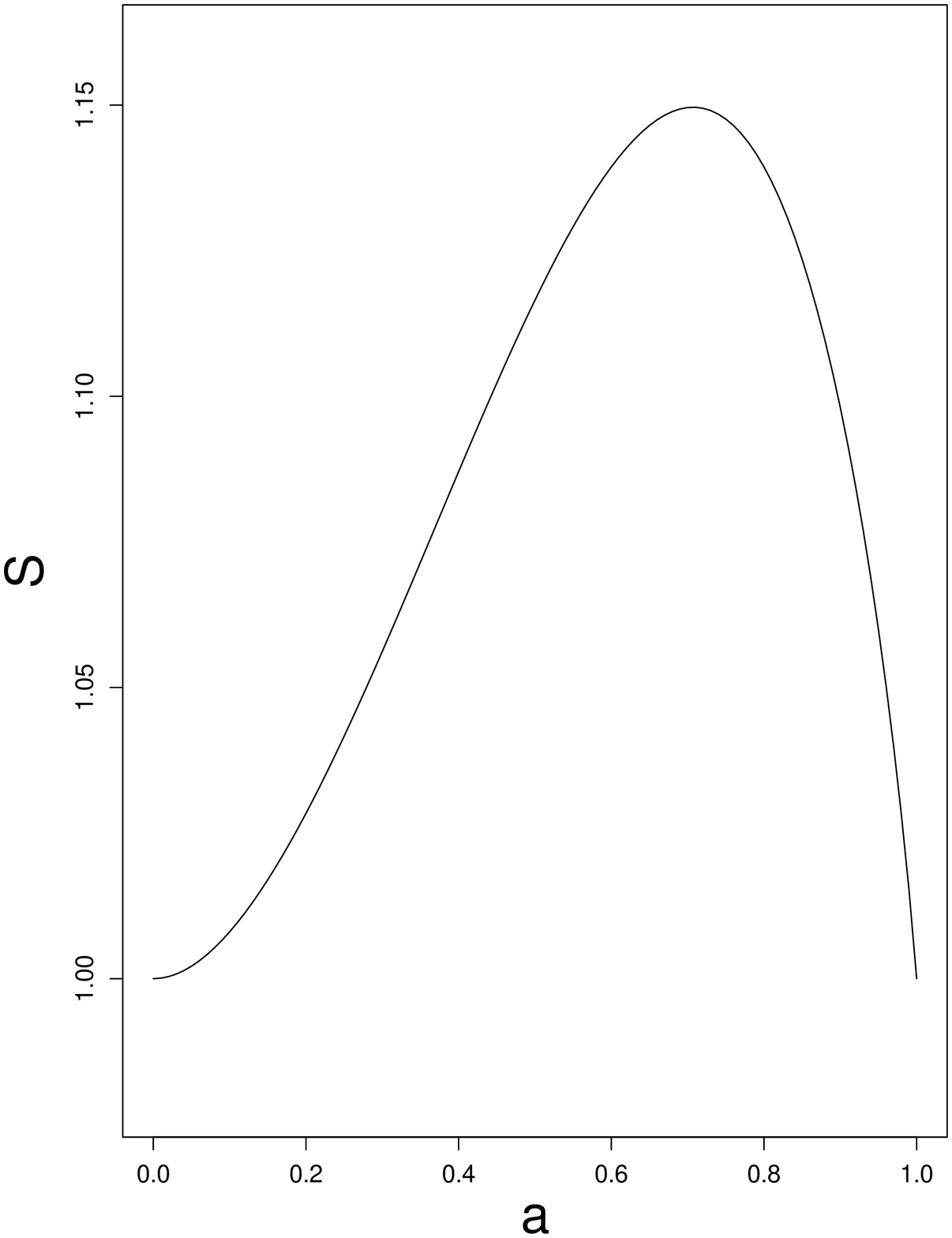}
\end{tabular}
\caption{The left-side graphic is the same as the left-side part of Fig. 2, but only representing the central peak and considering and additional blue curve, $a=0.7$. The right-side graphic is as the left-side one in Fig. 3 for the values of $\sigma$ ( the parameter of the mode distribution, Eq. (4)) corresponding to the first graphic, that is, $\sigma =\xi =1$ and $\bar{\sigma}= \bar{\xi}=6.$ }
\end{figure}
\section{Discussion}

In this paper we have considered a new example of multi-particle interference. The arrangement is similar to that in \cite{yjp}, but using distinguishable particles instead of identical ones. Then, we can analyse entanglement-based effects in the interference pattern instead of identity-induced ones. The multi-particle superposition leads to modifications of the patterns of simultaneous detection that can be used to study some aspects of multi-particle interference previously not considered in the literature.  

We have compared in Fig. 2 three different patterns. The comparison between these of superposition and mixed states informs about multi-particle coherence. In Fig. 2 right we see that the coherence leads to large differences in the  values of the respective peaks. On the other hand, comparing the mixture and product state patterns (using the two figures) we infer that the presence of two multi-particle terms, although they are not coherent, also modifies the product state interference (the mixed state is entangled). This way we can independently verify that the two elements generating entanglement contribute to the modifications of the patterns. Finally, comparing superposition and product states we can observe the entanglement effects when both elements are simultaneously taken into account. We have that in our example both effects do not add but tend to cancel each other.  

The large differences between the relative patterns represented in Fig. 2 can be used to verify multi-particle superposition. The small difference between superposition and product state interference makes very difficult to experimentally settle by comparison of the patterns if a state is in a multi-particle superposition . However, we can address the question in a two-step way. First, we obtain the mixture pattern. We repeat the interference experiment $n|a|^2$ times preparing the initial state in the product one giving $\Phi _a$, and $n|b|^2$ times in that leading to $\Phi _b$ (with $n$ the number of repetitions of the experiment). Second, we obtain the superposition pattern. Then, comparing the mixture pattern with these of product states (for instance, with that of $\Phi _a$, which has already been obtained in the first round of the experiment) and superposition we can verify (now the differences of the relative patterns are large) that there are several terms and that they are coherent. Consequently, we can verify the presence of a multi-particle superposition. For other values of the parameters the differences between the relative patterns are not so large and the scheme cannot be applied.     

We have also seen that the entanglement-induced effects do not show a monotone behaviour with respect to the Schmidt number of the initial state. This result is interesting because it strongly suggests that the dynamics of a multi-particle system under non-separability conditions is very complex, and cannot be described by a single magnitude as the entanglement degree. Even a qualitative concept as the approximate separation between the superposed and non superposed states does not show a simple trend with respect to the Schmidt number.  

A topic not treated here is the complementarity between one- and two-particle interference. As it is well-known theoretically \cite{Ze1}, and has been experimentally demonstrated for light \cite{2}, the two-particle visibility increases with entanglement, whereas the one-particle one decreases. In our case, as the entanglement depends on the coefficients of the superposition and the non-orthogonality of the one-particle states, we expect a dependence of the two visibilities on the overlapping of the non-orthogonal states. For fixed coefficients of the superposition the two-particle visibility should diminish with the increase of the one-particle overlapping, disappearing in the limit of full overlapping. The analysis of the complementarity relations in our arrangement would deserve further attention.  

It must be noted that the entanglement degree is not conserved during the evolution of the system. As signalled before some particles can be absorbed leading to a non-unitary evolution. In the path integral formalism used here the absorption is equivalent to the removal of some paths. The arrangement is outside the scope of the Local Operations Classical Communication (LOCC) paradigm \cite{ple}. In effect, when only one particle of a pair is absorbed, the path of the other particle does not contribute (we post-select the pairs where both particles are detected). This is a non-local operation where a (non-unitary) action on a particle affects to the two members of the pair. Outside LOCC, the Schmidt number of the system after the interaction with the diffraction grating can be different from the initial one. It would be interesting to evaluate the modified value of the entanglement degree and to see if there is a monotone behaviour with respect to it. 

\section*{Appendix 1: Wave functions}

We adapt the wave functions derived in \cite{yjp} to the case of distinguishable particles with masses $m_1$ and $m_2$. The two slits are denoted $A$ and $B$. After the slits the not normalized one-particle wave functions are $\psi _A+\psi _B$, ... Their explicit form is
\begin{equation}
\psi _A(x,t)= {\cal C}_1(\sigma)e^{im_1x^2/2\hbar (t-t_s)}e^{-(\alpha_1(\sigma) -i\beta_1(\sigma) )x^2} e^{-(\delta_1(\sigma)+i\gamma _1(\sigma))x}
\end{equation} 
and
\begin{equation}
\psi _B(x,t)= {\cal C}_1(\sigma)e^{im_1x^2/2\hbar (t-t_s)}e^{-(\alpha_1(\sigma) -i\beta_1(\sigma) )x^2} e^{(\delta_1(\sigma)+i\gamma _1(\sigma))x}
\end{equation} 
with $t_s$ the time at which the particles reach the slit,
\begin{eqnarray}
{\cal C}_1(\sigma)=\pi ^{-1/4} \left( \frac{1}{\sigma} + \frac{i\hbar \sigma t_s}{m_1}  \right) ^{-1/2} \left( \frac{m_1}{2i\hbar (t-t_s)(D_1+iF_1)}  \right) ^{1/2} \times \nonumber \\
e^{-x_0^2/2b_s^2}e^{G^2(D_1(\sigma)-iF_1(\sigma))/4(D_1^2(\sigma)+F_1^2(\sigma))}
\end{eqnarray}
\begin{equation}
\alpha_1(\sigma)=\frac{D_1(\sigma)H_1^2}{4(D_1^2(\sigma)+F_1^2(\sigma))};\beta_1(\sigma)=\frac{F_1(\sigma)H_1^2}{4(D_1^2(\sigma)+F_1^2(\sigma))}
\end{equation}
\begin{equation}
\gamma_1(\sigma)=\frac{D_1(\sigma)GH_1}{2(D_1^2(\sigma)+F_1^2(\sigma))};\delta_1(\sigma)=\frac{GH_1F_1(\sigma)}{2(D_1^2(\sigma)+F_1^2(\sigma))}
\end{equation}
\begin{equation}
D_1(\sigma)=\frac{1}{2b_s^2}+\frac{\sigma^2}{\mu _1(\sigma)}; F_1(\sigma)=-\frac{\hbar \sigma^4t_s}{m_1\mu _1(\sigma)}-\frac{m_1}{2\hbar (t-t_s)}
\end{equation}
and
\begin{equation}
G=\frac{x_0}{b_s^2}; H_1=\frac{m_1}{\hbar (t-t_s)}; \mu_1(\sigma)=2\left( 1+\frac{\hbar^2 \sigma^4 t^2}{m_1^2} \right)
\end{equation}
The rest of the one-particle wave functions can be obtained in the same way, introducing in the equations $m_1$ or $m_2$ and $\sigma$, $\bar{\sigma}$, $\chi$ or $\bar{\chi}$ according to the case considered.

\section*{Appendix 2: Dependence on the fixed detector}

We analyse in this Appendix the dependence of the joint patterns on the choice of the fixed detection position. We vary it from $y=0$ to all the range of possible values.  In Fig. 5 we include three of these curves. 
\begin{figure}[H]
\centering
\includegraphics[width=8cm,height=5cm]{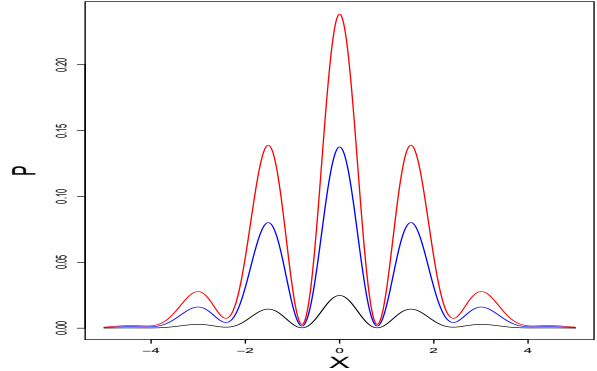}
\caption{Representation of $P$ versus $x$ as in Fig. 2 left, for the fixed points $y=0,0.7,1.7$ (red, black, blue).}
\end{figure}
The shape of the curves is very similar but the intensity of the maxima strongly depends on the value of the fixed point. For $y=0$ we have the larger values. When $y$ increases the values of the peaks sharply decrease ($y=0.7$), reaching a minimum around $y=0.9$. For higher values of $y$ the behaviour changes from a decrease to an increase of the values of the maxima. Around $y=1.7$ we reach a relative maximum of the peak values. We observe an oscillation of the maxima values of the patterns when the fixed point changes. The same oscillatory behaviour persists for larger values of $y$, being the peaks values progressively smaller. The same comportment is observed for other values of the degree of non-orthogonality (without modifying the coefficients of the superposition).       

We must discuss the connection of these oscillations with the concept of conditional interference, introduced in \cite{Ze1} and experimentally studied for light in \cite{bra}. A pattern shows conditionality when the location of the fringes depends in an oscillatory way on the relative positions of the detectors. As discussed in detail in \cite{bra} the interference patterns are in general the sum of conditional and non-conditional contributions, and only for some input light states the conditionality becomes manifest. In our case we do not observe conditional behaviour. The non-conditional terms are dominant. The prevalence of the last contributions can be easily understood. The relation between both types of contribution reflects a sort of complementarity between one- and two-particle interference \cite{Ze1}.  As shown in Fig. 2 left, the superposition pattern is very close to that of the product state, indicating that the multi-particle contributions, and consequently the conditional ones, are small when compared to the non-conditional ones. The impossibility of observing conditionality does not imply the absence of conditional terms and, most important, the possibility of determining other multi-particle effects (as for example the modifications with respect to product state patterns) . It would be interesting, following the approach in \cite{bra}, to find input states where the conditionality could be observed.

\end{document}